\documentclass[preprint, amsmath,amssymb,prl,superscriptaddress]{revtex4-1}
\usepackage{graphicx}% Include figure files
\usepackage{dcolumn}% Align table columns on decimal point
\usepackage{bm}% bold math
\usepackage{epsfig}
\usepackage{natbib}
\usepackage{color}
\usepackage[normalem]{ulem}
\usepackage{enumerate}

\newcommand{\Beff}{B_\text{eff}} %electron free mass
\newcommand{\mCF}{m_\text{CF}} %composite fermion mass
\newcommand{\lB}{l_\text{B}} % magnetic length
\newcommand{\Rxx}{R_\text{xx}}  % Rxx
\begin{document}

\title{Phase transition from a composite fermion liquid to a Wigner solid in the lowest Landau level of ZnO}
\author{D.~Maryenko}
\email{maryenko@riken.jp}
\affiliation{RIKEN Center for Emergent Matter Science (CEMS), Wako 351-0198, Japan}
\author{A.~McCollam}
\affiliation{High Field Magnet Laboratory (HFML-EMFL) and Institute for Molecules and Materials, Radboud University, 6525 ED Nijmegen, The Netherlands}
\author{J.~Falson}
\affiliation{Department of Applied Physics and Quantum-Phase Electronics Center (QPEC), The University of Tokyo, Tokyo 113-8656, Japan}
\author{Y.~Kozuka}
\affiliation{Department of Applied Physics and Quantum-Phase Electronics Center (QPEC), The University of Tokyo, Tokyo 113-8656, Japan}
\author{J.~Bruin}
\affiliation{High Field Magnet Laboratory (HFML-EMFL) and Institute for Molecules and Materials, Radboud University, 6525 ED Nijmegen, The Netherlands}
\author{U.~Zeitler}
\affiliation{High Field Magnet Laboratory (HFML-EMFL) and Institute for Molecules and Materials, Radboud University, 6525 ED Nijmegen, The Netherlands}
\author{M.~Kawasaki}
\affiliation{RIKEN Center for Emergent Matter Science (CEMS), Wako 351-0198, Japan}
\affiliation{Department of Applied Physics and Quantum-Phase Electronics Center (QPEC), The University of Tokyo, Tokyo 113-8656, Japan}

%\today

\begin{abstract}
\textbf{Interactions between the constituents of a condensed matter system can drive it through a plethora of different phases due to many-body effects. 
A prominent platform for this type of behavior is a two-dimensional electron system in a magnetic field, which evolves intricately through various gaseous, liquid and solids phases governed by Coulomb interaction~\cite{perspective, JainCF, HLRTheory}.
Here we report on the experimental observation of a phase transition between the Laughlin liquid of composite fermions and the adjacent insulating phase of a magnetic field-induced Wigner solid~\cite{Laughlin,WSGaAsHoles}. 
The experiments are performed in the lowest Landau level of a MgZnO/ZnO two-dimensional electron system with attributes of both a liquid and a solid~\cite{MaryenkoPRL, Falson2015}. 
An in-plane magnetic field component applied on top of the perpendicular magnetic field extends the Wigner phase further into the liquid phase region. 
Our observations indicate the direct competition between a Wigner solid and a Laughlin liquid both formed by composite particles rather than bare electrons.}
%\textbf{
%Interactions between its constituents can drive a system through a plethora of different phases governed by many-body effects. 
%A prominent platform to such is a two-dimensional electron system in a magnetic field that evolves intricately through various gaseous, liquid and solids phases governed by Coulomb interaction~\cite{perspective, JainCF, HLRTheory}.
%Here we report on the experimental observation of a phase transition between the Laughlin liquid of composite fermions and the insulating phase of a magnetic field-induced
%Wigner solid adjacent to it~\cite{Laughlin,WSGaAsHoles}. 
%The experiments are performed in the lowest Landau level of a MgZnO/ZnO two-dimensional electron system with attributes of both a liquid and a solid~\cite{MaryenkoPRL, Falson2015}. 
%An in-plane field component applied on top of the perpendicular magnetic field extends the Wigner phase further into the liquid phase field region. 
%Our observations indicate the direct competition between a Wigner solid and a Laughlin liquid both formed by composite particles rather than bare electrons.}
\end{abstract}
\maketitle

\begin{figure}%[p]
\includegraphics{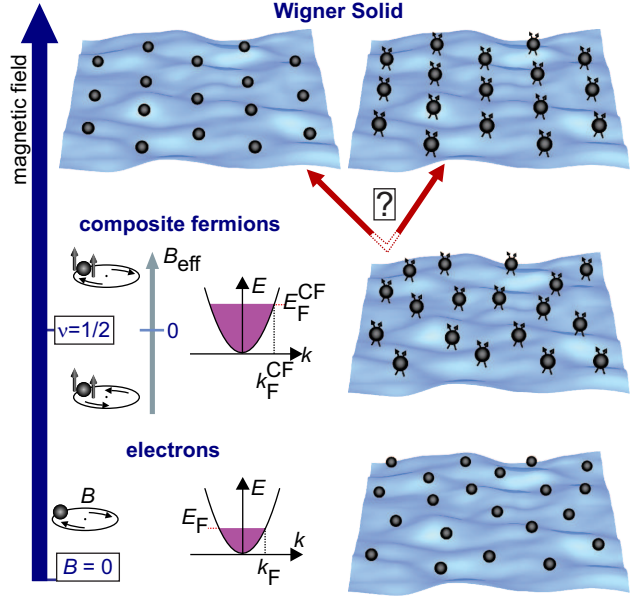}
\caption{\label{Introduction} \textbf{Concept Figure:} 
The different phases of a two-dimensional electron system (2DES)  in a magnetic field. At zero magnetic field (bottom panel) the electrons are described as a weakly interacting Fermi gas with a well-defined Fermi surface.  In the half-filled lowest LL, e.g at filling factor $\nu=$1/2, the electrons reduce their mutual interaction by attaching the two magnetic flux quanta, resulting in the emergence of new particles, so-called composite fermions (middle panel)~\cite{HLRTheory, JainCF}. These particles form a Fermi surface at $\nu=$1/2 and move in an effective field $B_\text{eff}=B-B_{\nu=1/2}$  giving rise to magnetoresistance oscillations known as the fractional quantum Hall effect (Figure~\ref{FigMain}).  At even lower filling factors, a Wigner solid, a crystalline phase of electrons arranged by the repulsive Coulomb force and another manifestation of many-body correlations, becomes the ground state, which can be formed either by bare electrons (top left)  or composite fermions (top right). }
\end{figure}

A magnetic field $B$ applied perpendicularly to a two-dimensional  system modifies its density of states by arranging the charge carriers in discrete Landau levels (LLs). 
Additionally, the Coulomb interaction acting on the magnetic length scale $\lB=\sqrt{\hbar/e B}$ can be tuned by the magnetic field, causing the high mobility carriers to evolve through various correlated phases~\cite{perspective} , see Fig.~\ref{Introduction}.

In the lowest LL the Wigner phase competes with the liquid phase and manifests as a large magnetoresistance peak  around or below $\nu=1/3$~\cite{WSGaAsHoles, KozukaInsulating}. Thus the electron system, when moving from $\nu=$1/2 to lower $\nu$'s, has to undergo a reorganization of the ground state between the picture of  composite fermions describing the magnetoresistance oscillations around $\nu=1/2$ and the bare electrons forming the Wigner solid at smaller $\nu$'s. An alternative concept for such a regime is the realization of a Wigner solid formed by the composite fermions  (Fig.~\ref{Introduction}, top right) - an idea that was put forward in a number of theoretical works~\cite{ WignerSolidCFTheory1, WignerSolidCFTheory2, WignerSolidCFTheory3, WignerSolidCFTheory4}.    Recent experiments focusing on GaAs-based 2DES have been gradually accumulating evidences pointing towards the realization of CF Wigner solid~\cite{Chen2004, WignerSolidCFExperiment, Shayegan2014, Zhang2015, Ashoori2016}. Here, we study the magnetotransport in a ZnO heterostructure (see Methods)  in the magnetic field region between the CF liquid phase formed at $\nu=1/2$ and the insulating phase appearing at higher field, associated with the Wigner phase. The mangetotransport features in this region exhibit a character of both CF liquid and Wigner solid. The presence of such a region  with this interlaced character has a number of plausible explanations, one of which is  the presence of the Wigner phase formed by composite fermions.

\begin{figure*}%[p]
\includegraphics{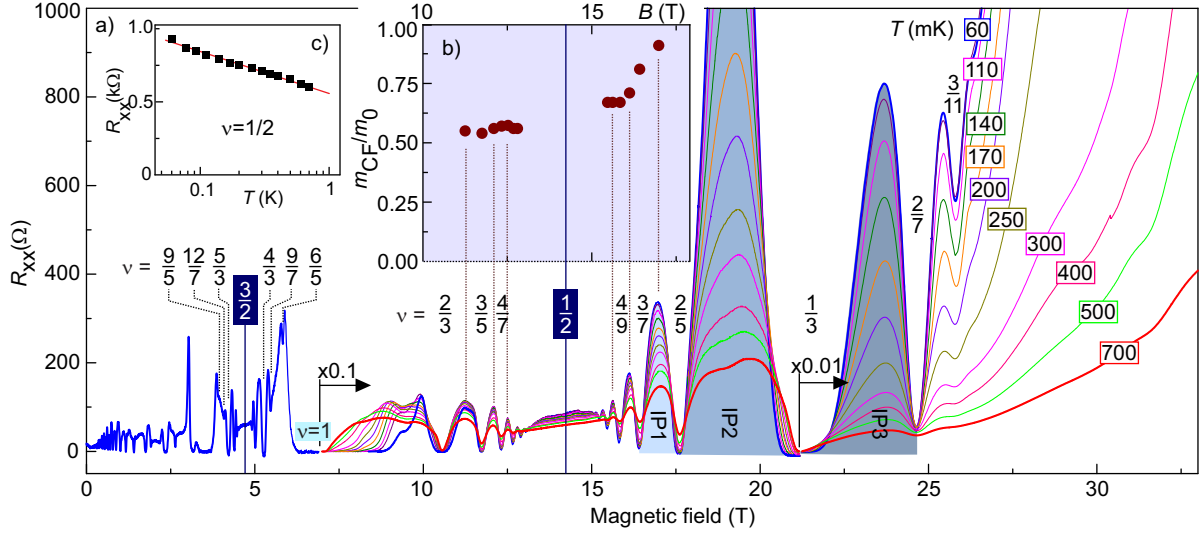}
\caption{\label{FigMain} \textbf{Temperature-dependent magnetotransport  up to 33~T}. a) $\Rxx$ at $T = 60$~mK (blue trace). The other colours show $\Rxx (B)$ in the fractional quantum Hall regime for higher temperatures. The insulating phases IP1, IP2 and IP3 associated with the Wigner solid are indicated by the shaded region. b) Mass of composite fermions extracted from the temperature dependence of the $\Rxx$ oscillation amplitude. c) Temperature dependence of $\Rxx$ at $\nu = 1/2$. The decreasing resistance with increasing temperatures indicates a residual interaction between composite fermions.}
\end{figure*}

Figure~\ref{FigMain} shows a full scan of the magnetotransport from 0~T to 33~T applied perpendicular to the 2DES plane. Several fractional quantum Hall states are observed around $\nu = 3/2$, consistent with  previous results~\cite{Falson2015} and, in addition, developing minima are observed at $\nu = $9/5, 12/7, 9/7, and 6/5. Furthermore, up to six fractional quantum Hall states are observed on both sides of $\nu = 1/2$.  
Close inspection of the transport around $\nu$ = 1/2 reveals a distinct asymmetry; $\Rxx$ maxima between fractional quantum Hall states for $\nu < 1/2$ are much larger than those for $\nu > 1/2$. This increase becomes increasingly dramatic between $\nu$ = 2/5, 1/3 and 2/7.  Such a high resistance phase between $\nu=$1/3 and 2/7 has also been observed in GaAs,  and was interpreted as the electron Wigner solid pinned by disorder. Two mechanisms have been identified for the appearance of the Wigner solid around these filling factors: one is the Landau level mixing, which modifies the ground state energies of fractional quantum Hall states and Wigner solid~\cite{WSGaAsHoles}; the other is short range disorder~\cite{WSGaAsHoles, WSDisorder, WSPinningDisorder}.  Both mechanisms are distinctively more pronounced in ZnO-heterostructures than in GaAs~\cite{MaryenkoPRL, ShortRangeZnO}, and the ZnO system is, therefore, ideal to access the competition between liquid and solid phases in the fractional quantum Hall regime.

 In Fig.~\ref{FigMain} the distinct high resistance phases are colored and marked  as  IP1, IP2 and IP3. On the basis of Wigner solid studies in other materials system, the characteristics of IP1, IP2 and IP3 are typical of the Wigner solid. The temperature dependence of IP1, IP2 and IP3 resembles the melting of the Wigner solid (Supplementary Fig.~1a). The non-linear current-voltage characteristics are associated with the depinning of the Wigner solid from the disorder,  when a certain threshold force is exceeded, and its subsequent sliding along the disorder landscape (Supplementary Fig.1b).   Thus, the trace of a Wigner solid appears  already in IP1 between $\nu=3/7$ and $\nu=2/5$,  whereas a larger $\Rxx$ and  \textit{I-V} non-linearity at IP2 and IP3 indicate an even more pronounced Wigner solid.

IP1 represents an interesting region. While it shows the features of an emerging Wigner solid, it is at the same time a part of  the $\Rxx$ oscillations caused by the composite fermions' orbital motion in $\Beff$, and therefore can also be attributed to the liquid phase. Therefore, we now analyze the CF mass $\mCF$ around $\nu=1/2$ from the temperature dependence of  the$\Rxx$ oscillation amplitude by using the Lifshitz-Kosevitch formalism (Supplementary Information).  Figure~\ref{FigMain}b displays $\mCF$ around $\nu=1/2$, which extends the linear dependence of $\mCF$ on $B$ to higher field~\cite{MaryenkoPRL}.  More noticeable is the excessive increase of $\mCF$ over the linear trend when the 
2DES approaches the insulating phase IP1. The mass increase can be interpreted as a signature of the underlying particles becoming more inert due to 
the formation of a solid phase. This is then in agreement with observing the traces of the Wigner solid character at IP1, and more strongly pronounced Wigner phases IP2 and IP3 at higher field. Since we cannot assume the coexistence of electrons and composite fermions, as it would require a simultaneous existence of two gauge fields, the dual character of IP1 and IP2 has to be consolidated within the model frame based on either electrons or composite fermions. Because of the multiple experimental evidences for the validity of CF picture~\cite{perspective}, it would be more natural to treat the transport anomaly in ZnO around $\nu=1/2$ with the composite fermion as an underlying particle for both liquid and solid phases. Another factor favoring the Wigner solid formation from composite fermions is the residual interaction among composite fermions. Indeed, the logarithmic temperature dependence of $\Rxx$ at $\nu=1/2$ shown in Fig.~\ref{FigMain}c points towards a residual CF interaction~\cite{HLRTheory, MaryenkoPRL, Rokhinson1995}.

\begin{figure*}%[p]
\includegraphics{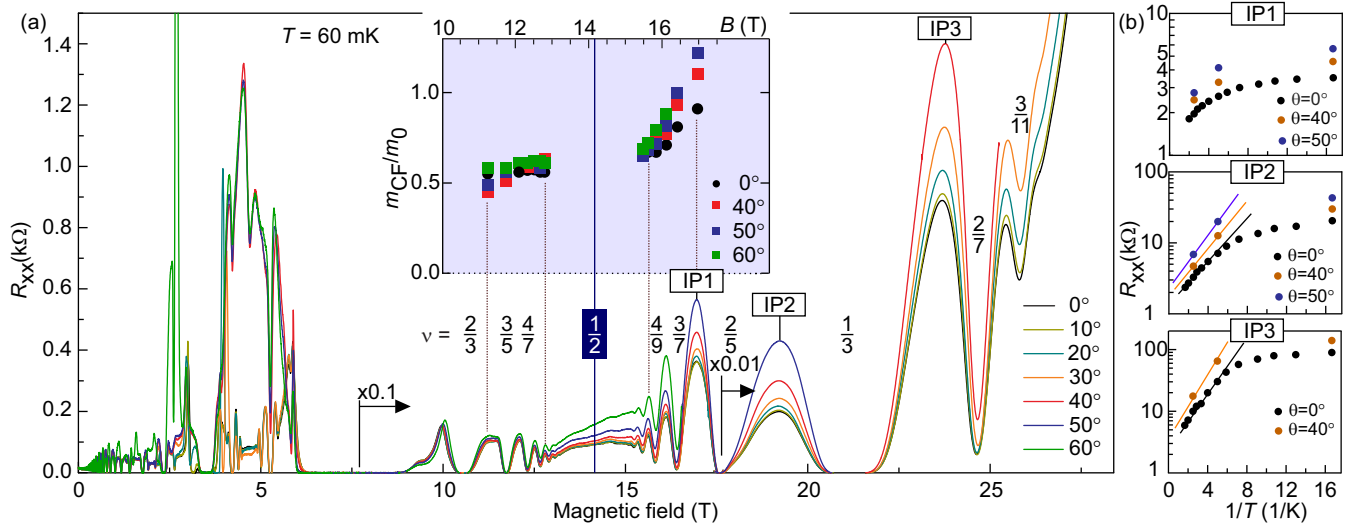}
\caption{\label{FigRotation} \textbf{Magnetotransport in tilted magnetic fields}. (a) The resistance of insulating phases IP1, IP2 and IP3 increases with the application of in-plane magnetic field. The insulating phase shifts towards higher filling factors with increasing in-plane field, as can be seen from the growing background around $\nu=1/2$.  Inset: mass of composite fermions evaluated from temperature dependence of $\Rxx$ oscillation amplitude. (b) Temperature dependence of insulating phases IP1, IP2 and IP3 at several tilt angles.
\vspace*{10cm}}
\end{figure*}

The transport properties discussed above change dramatically when the sample is rotated in the magnetic field, that is, when an additional field component is applied parallel to the 2DES. 
Since the electron spin susceptibility for this structure is about 2, the opposite spin orientation branch of the lowest LL lies energetically high and is not populated. Thus the spin effects are not anticipated to play a role for the discussion below. 
Figure~\ref{FigRotation}a depicts $\Rxx$ traces at several sample orientations $\theta$ obtained at  base temperature and shows the asymmetrical impact of the in-plane field on the transport for $\nu<1/2$ and $\nu>1/2$ ($\theta$ is the tilt angle between the normal of the 2DES plane and the magnetic field direction). Firstly, one notices that $\Rxx$ of  IP1, IP2 and IP3 increases gradually with an increasing $\theta$, while $\Rxx$ minima at $\nu=$ 3/7, 2/5 and 1/3 do not change significantly.  
Thus the Wigner phase becomes more pronounced by applying an in-plane field. The temperature dependence of the insulating phases is depicted for three representative $\theta$s in Fig.~\ref{FigRotation}b.  
Furthermore, $\Rxx$  around $\nu=1/2$ gains a background, which becomes larger with the increasing $\theta$.
Since magnetotransport experiments in GaAs demonstrate the extension of the tail of the insulating phase into the $\nu=1/2$ region with an increasing $\theta$~\cite{PRBTiltFieldGaAs,3DWigner}, we may also suppose that the background forming around $\nu=1/2$ has the same origin and is associated with the insulating phase shifting towards $\nu=1/2$ and above.

 It is noteworthy that the $\Rxx$ oscillations are not damped but rather persist on top of the background. We analyze the temperature dependence of the $\Rxx$ oscillation amplitude and estimate $\mCF$ around $\nu=1/2$ for several $\theta$s (Supplementary Information).
The inset of Fig.~\ref{FigRotation} depicts the result of this analysis.  For $\nu>1/2$, $\mCF$ does not show any noticeable change, but it shows a pronounced field and tilt angle dependence for $\nu<1/2$, that is, for a given perpendicular magnetic field  $\mCF$ is heavier at a larger tilt angle. The mass increase serves as a sign of the CFs becoming more strongly localized. This is consistent with the growing insulating character of IP1 and the shift of the Wigner phase towards higher $\nu$'s with  increasing $\theta$.

\begin{figure}[!thb]%[p]
\includegraphics{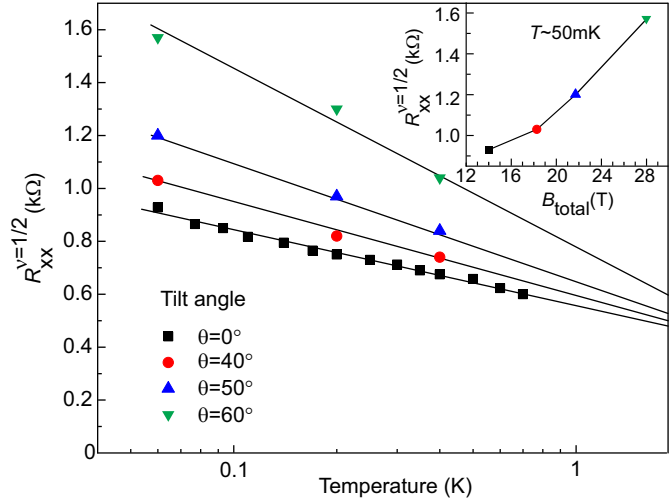}
\caption{\label{Fig4} \textbf{Temperature dependence of $\Rxx$ at filling factor $\nu=$1/2 for different tilt angles.} The resistance decreases with increasing temperature indicating a residual interaction between the composite fermions at zero tilt angle. The slope becomes more pronounced at higher tilt angles, i.e. a stronger in-plane field, and points towards a more robust Wigner solid phase.}
\end{figure}

Finally, the enhanced CF interaction with an added in-plane field also becomes apparent at $\nu=1/2$: Figure~\ref{Fig4} presents the temperature dependence $\Rxx$ at several $\theta$'s and shows that the slope of the logarithmic temperature dependence increases with $\theta$. The slope at each $\theta$ reflects not only the CF residual interaction but also the melting of Wigner phase penetrating to higher filling factors with increasing $\theta$. 

In order to further address this in-plane field induced stabilization of the Wigner solid we now analyse how much the in-plane field squeezes the electron wave function, as it effectively enhances the Coulomb interaction and can affect the transport properties~\cite{Feng, CFInTiltedField, DasSarmaWaveFunctionWidth}. In zero in-plane field the wave function of the heterostructure is about 10~nm wide. 
At $\theta=50^\circ$ it is squeezed down to 2.6~nm at $B_\bot=12$~T, representing the region $\nu>1/2$, and down to 2.2~nm at $B_\bot=17$~T, representing the region $\nu<1/2$ (Supplementary Information). 
Since the wave function width reduces significantly with in-plane field on both sides of $\nu=$1/2 compared with zero in-plane field, the Coulomb interaction  should also be equally enhanced around $\nu=1/2$. 
Nonetheless, $\mCF$ defined by the interaction effects remains almost unchanged  for $\nu>1/2$ and no large effect of in-plane field on transport characteristics is seen in this region. 
Consequently, the increase of $\mCF$ for $\nu<1/2$ is not mainly caused by the reduced wave function width. 
Rather, it supports our hypothesis of a Wigner solid formation. Since the solid phase gains over the liquid phase upon the application of the in-plane field, $\mCF$ increase  in $\nu<1/2$ region reflects an effective localization of the composite fermions. The origin for the asymmetrical response of liquid and solid phases to the in-plane field remains an open question, but our experimental result can likely be the precursor for the new insulating state proposed by Piot et al. \cite{3DWigner}.

Our experimental data show that the electron system enters an unconventional correlation regime, which reflects the character of both solid and liquid phases for $\nu<1/2$.  One interpretation for such regime can be the formation (melting) of the Wigner solid upon increasing (decreasing) the magnetic field, where a composite fermion would be the underlying particle in both phases. In such a regime, the particles can form a hexatic phase characterized by bond-oriented nearest-neighbor ordering, while the phase transition obeys the Kosterlitz-Thouless model~\cite{HalperinNelson1978, Morf1979, Young1979,  Strandburg1988}. In another scenario, a transition between the liquid and solid phase in a two-dimensional system can be accompanied by the appearance of  intermediate phases, such as microemulsion phases associated with liquid crystalline phases~\cite{ME1-PRB2004, ME3-RMP2010, Strandburg1988}.  
It appears unlikely that the magnetic field regime with dual character can be modeled by assuming the co-existence of  electrons and composite fermions, as it would then require a simultaneous existence of two gauge fields. Our experimental results are interpreted within the composite fermions approach, which has recently attracted renewed attention from  theory predicting that the composite fermions can be Dirac particles ~\cite{CFModel1, CFModel2, CFModel3, CFModel4}. This also introduces an exciting perspective for ZnO studies. Our experimental results sheds the light on the composite fermion paradigm in a system distinct from conventional semiconductors and also on how the charge carrier system translates between liquid and solid phases.
 
\vspace*{2em}
\noindent
\textbf{Methods}\\
\textbf{Sample}  The sample under study is a MgZnO/ZnO heterostructure with a charge carrier density $n=1.7\times10^{11}$~cm$^{-2}$ and a mobility $\mu=600,000$~cm$^2$/Vs at the base temperature of our dilution refrigerator $T = 60~$ mK.\\ 
\textbf{Tilted-field magnetotransport} The sample is mounted on a rotating stage allowing \textsl{in-situ} sample rotation in the magnetic field. The tilt angle $\theta$ is determined acurately from the shift of $\Rxx$ resistance minima of the well-known fractional quantum Hall states.

\vspace*{2em}
\noindent
\textbf{Acknowledgment} We acknowledge the support of the HFML-RU/FOM member of the European Magnetic Field Laboratory (EMFL). We would like to thank M. Kawamura, A. S. Mishchenko, M. Ueda, K. von Klitzing  and N. Nagaosa for fruitful discussion. 

\vspace*{2em}
\noindent
\textbf{Author Conrinbutions}  M.K. initiated and supervised the project. D.M. concieved and designed the experiment. J.F. and Y.K. fabricated the samples. D.M., A.M., J.B. and U.Z. performed the high-field magnetotransport experiment and analysed the data. D.M. and U.Z.  wrote the paper with input from all co-authors.


\begin{thebibliography}{10}
\expandafter\ifx\csname url\endcsname\relax
  \def\url#1{\texttt{#1}}\fi
\expandafter\ifx\csname urlprefix\endcsname\relax\def\urlprefix{URL }\fi
\providecommand{\bibinfo}[2]{#2}
\providecommand{\eprint}[2][]{\url{#2}}

\bibitem{perspective}
\bibinfo{author}{Das~Sarma, S.} \& \bibinfo{author}{Pinczuk, A.}
\newblock \emph{\bibinfo{title}{Perspective in quantum Hall effects}}
  (\bibinfo{publisher}{John Wiley}, \bibinfo{year}{1997}).

\bibitem{HLRTheory}
\bibinfo{author}{Halperin, B.~I.}, \bibinfo{author}{Lee, P.~A.} \&
  \bibinfo{author}{Read, N.}
\newblock \bibinfo{title}{Theory of the half-filled Landau level}.
\newblock \emph{\bibinfo{journal}{Phys. Rev. B}} \textbf{\bibinfo{volume}{47}},
  \bibinfo{pages}{7312--7343} (\bibinfo{year}{1993}).

\bibitem{JainCF}
\bibinfo{author}{Jain, J.~K.}
\newblock \bibinfo{title}{Composite-fermion approach for the fractional quantum Hall effect}.
\newblock \emph{\bibinfo{journal}{Phys. Rev. Lett.}}
  \textbf{\bibinfo{volume}{63}}, \bibinfo{pages}{199--202}
  (\bibinfo{year}{1989}).
  
  \bibitem{Laughlin}
\bibinfo{author}{Laughlin, R.~B.}
\newblock \bibinfo{title}{Anomalous quantum Hall effect: an incompressible quantum fluid with fractionally charged excitations}.
\newblock \emph{\bibinfo{journal}{Phys. Rev. Lett.}}
  \textbf{\bibinfo{volume}{50}}, \bibinfo{pages}{1395}
  (\bibinfo{year}{1983}).

\bibitem{WSGaAsHoles}
\bibinfo{author}{Santos, M.~B.} \emph{et~al.}
\newblock \bibinfo{title}{Observation of a reentrant insulating phase near the 1/3 fractional quantum hall liquid in a two-dimensional hole system}.
\newblock \emph{\bibinfo{journal}{Phys. Rev. Lett.}}
  \textbf{\bibinfo{volume}{68}}, \bibinfo{pages}{1188--1191}
  (\bibinfo{year}{1992}).
  
\bibitem{MaryenkoPRL}
\bibinfo{author}{Maryenko, D.} \emph{et~al.}
\newblock \bibinfo{title}{Temperature-dependent magnetotransport around $\nu$=1/2 in ZnO heterostructures}.
\newblock \emph{\bibinfo{journal}{Phys. Rev. Lett.}}
  \textbf{\bibinfo{volume}{108}}, \bibinfo{pages}{186803}
  (\bibinfo{year}{2012}).

\bibitem{Falson2015}
\bibinfo{author}{Falson, J.} \emph{et~al.}
\newblock \bibinfo{title}{Even-denominator fractional quantum Hall physics in ZnO}.
\newblock \emph{\bibinfo{journal}{Nat. Physics}} \textbf{\bibinfo{volume}{11}},
  \bibinfo{pages}{347} (\bibinfo{year}{2015}).


\bibitem{KozukaInsulating}
\bibinfo{author}{Kozuka, Y.} \emph{et~al.}
\newblock \bibinfo{title}{Insulating phase of a two-dimensional electron gas in  Mg${}_{x}$Zn${}_{1\ensuremath{-}x}$O/ZnO heterostructures below
  $\ensuremath{\nu}=\frac{1}{3}$}.
\newblock \emph{\bibinfo{journal}{Phys. Rev. B}} \textbf{\bibinfo{volume}{84}},
  \bibinfo{pages}{033304} (\bibinfo{year}{2011}).

\bibitem{WignerSolidCFTheory1}
\bibinfo{author}{Archer, A.~C.} \& \bibinfo{author}{Jain, J.~K.}
\newblock \bibinfo{title}{Static and dynamic properties of type-II composite  fermion Wigner crystals}.
\newblock \emph{\bibinfo{journal}{Phys. Rev. B}} \textbf{\bibinfo{volume}{84}},
  \bibinfo{pages}{115139} (\bibinfo{year}{2011}).

\bibitem{WignerSolidCFTheory2}
\bibinfo{author}{Chang, C.-C.}, \bibinfo{author}{Jeon, G.~S.} \&
  \bibinfo{author}{Jain, J.~K.}
\newblock \bibinfo{title}{Microscopic verification of topological electron-vortex binding in the lowest Landau level crystal state}.
\newblock \emph{\bibinfo{journal}{Phys. Rev. Lett.}}
  \textbf{\bibinfo{volume}{94}}, \bibinfo{pages}{016809}
  (\bibinfo{year}{2005}).

\bibitem{WignerSolidCFTheory3}
\bibinfo{author}{Yi, H.} \& \bibinfo{author}{Fertig, H.~A.}
\newblock \bibinfo{title}{Laughlin-Jastrow-correlated Wigner crystal in a  strong magnetic field}.
\newblock \emph{\bibinfo{journal}{Phys. Rev. B}} \textbf{\bibinfo{volume}{58}},
  \bibinfo{pages}{4019--4027} (\bibinfo{year}{1998}).

\bibitem{WignerSolidCFTheory4}
\bibinfo{author}{Narevich, R.}, \bibinfo{author}{Murthy, G.} \&
  \bibinfo{author}{Fertig, H.~A.}
\newblock \bibinfo{title}{Hamiltonian theory of the composite-fermion Wigner crystal}.
\newblock \emph{\bibinfo{journal}{Phys. Rev. B}} \textbf{\bibinfo{volume}{64}},
  \bibinfo{pages}{245326} (\bibinfo{year}{2001}).

\bibitem{Chen2004}
\bibinfo{author}{Chen, Y.~P.} \emph{et~al.}
\newblock \bibinfo{title}{Evidence for two different solid phases of two-dimensional electrons in high magnetic fields}.
\newblock \emph{\bibinfo{journal}{Phys. Rev. Lett.}}
  \textbf{\bibinfo{volume}{93}}, \bibinfo{pages}{206805}
  (\bibinfo{year}{2004}).

\bibitem{WignerSolidCFExperiment}
\bibinfo{author}{Zhu, H.} \emph{et~al.}
\newblock \bibinfo{title}{Observation of a pinning mode in a Wigner solid with $\ensuremath{\nu}=1/3$ fractional quantum Hall excitations}.
\newblock \emph{\bibinfo{journal}{Phys. Rev. Lett.}}
  \textbf{\bibinfo{volume}{105}}, \bibinfo{pages}{126803}
  (\bibinfo{year}{2010}).

\bibitem{Shayegan2014}
\bibinfo{author}{Liu, Y.} \emph{et~al.}
\newblock \bibinfo{title}{Fractional quantum Hall effect and Wigner crystal of interacting composite fermions}.
\newblock \emph{\bibinfo{journal}{Phys. Rev. Lett.}}
  \textbf{\bibinfo{volume}{113}}, \bibinfo{pages}{246803}
  (\bibinfo{year}{2014}).

\bibitem{Zhang2015}
\bibinfo{author}{Zhang, C.}, \bibinfo{author}{Du, R.-R.},
  \bibinfo{author}{Manfra, M.~J.}, \bibinfo{author}{Pfeiffer, L.~N.} \&
  \bibinfo{author}{West, K.~W.}
\newblock \bibinfo{title}{Transport of a sliding Wigner crystal in the four flux composite fermion regime}.
\newblock \emph{\bibinfo{journal}{Phys. Rev. B}} \textbf{\bibinfo{volume}{92}},
  \bibinfo{pages}{075434} (\bibinfo{year}{2015}).

\bibitem{Ashoori2016}
\bibinfo{author}{Jang, J.}, \bibinfo{author}{Hunt, B.~M.},
  \bibinfo{author}{Pfeiffer, L.~N.}, \bibinfo{author}{West, K.~W.} \&
  \bibinfo{author}{Ashoori, R.~C.}
\newblock \bibinfo{title}{Sharp tunnelling resonance from the vibrations of an  electronic Wigner crystal}.
\newblock \emph{\bibinfo{journal}{Nature Physics}}
  \textbf{\bibinfo{volume}{13}}, \bibinfo{pages}{340--344}
  (\bibinfo{year}{2016}).

\bibitem{WSDisorder}
\bibinfo{author}{Li, W.}, \bibinfo{author}{Luhman, D.~R.},
  \bibinfo{author}{Tsui, D.~C.}, \bibinfo{author}{Pfeiffer, L.~N.} \&
  \bibinfo{author}{West, K.~W.}
\newblock \bibinfo{title}{Observation of reentrant phases induced by short-range disorder in the lowest Landau level of
  ${\mathrm{Al}}_{x}{\mathrm{Ga}}_{1\ensuremath{-}x}\mathrm{As}/{\mathrm{Al}}_{0.32}{\mathrm{Ga}}_{0.68}\mathrm{As}$
  heterostructures}.
\newblock \emph{\bibinfo{journal}{Phys. Rev. Lett.}}
  \textbf{\bibinfo{volume}{105}}, \bibinfo{pages}{076803}
  (\bibinfo{year}{2010}).

\bibitem{WSPinningDisorder}
\bibinfo{author}{Moon, B.-H.}, \bibinfo{author}{Engel, L.~W.},
  \bibinfo{author}{Tsui, D.~C.}, \bibinfo{author}{Pfeiffer, L.~N.} \&
  \bibinfo{author}{West, K.~W.}
\newblock \bibinfo{title}{Pinning modes of high-magnetic-field Wigner solids with controlled alloy disorder}.
\newblock \emph{\bibinfo{journal}{Phys. Rev. B}} \textbf{\bibinfo{volume}{89}},
  \bibinfo{pages}{075310} (\bibinfo{year}{2014}).


\bibitem{ShortRangeZnO}
\bibinfo{author}{K\"archer, D.~F.} \emph{et~al.}
\newblock \bibinfo{title}{Observation of microwave induced resistance and  photovoltage oscillations in MgZnO/ZnO heterostructures}.
\newblock \emph{\bibinfo{journal}{Phys. Rev. B}} \textbf{\bibinfo{volume}{93}},
  \bibinfo{pages}{041410} (\bibinfo{year}{2016}).

\bibitem{Rokhinson1995}
\bibinfo{author}{Rokhinson, L.~P.}, \bibinfo{author}{Su, B.} \&
  \bibinfo{author}{Goldman, V.~J.}
\newblock \bibinfo{title}{Logarithmic temperature dependence of conductivity at  half-integer filling factors: Evidence for interaction between composite
  fermions}.
\newblock \emph{\bibinfo{journal}{Phys. Rev. B}} \textbf{\bibinfo{volume}{52}},
  \bibinfo{pages}{R11588--R11590} (\bibinfo{year}{1995}).

\bibitem{PRBTiltFieldGaAs}
\bibinfo{author}{Pan, W.}, \bibinfo{author}{Cs\'athy, G.~A.},
  \bibinfo{author}{Tsui, D.~C.}, \bibinfo{author}{Pfeiffer, L.~N.} \&
  \bibinfo{author}{West, K.~W.}
\newblock \bibinfo{title}{Transition from a fractional quantum Hall liquid to an electron solid at Landau level filling $\ensuremath{\nu}=\frac{1}{3}$ in
  tilted magnetic fields}.
\newblock \emph{\bibinfo{journal}{Phys. Rev. B}} \textbf{\bibinfo{volume}{71}},
  \bibinfo{pages}{035302} (\bibinfo{year}{2005}).

\bibitem{3DWigner}
\bibinfo{author}{Piot, B.~A.} \emph{et~al.}
\newblock \bibinfo{title}{Wigner crystallization in a quasi-three-dimensional  electronic system}.
\newblock \emph{\bibinfo{journal}{Nature Physics}}
  \textbf{\bibinfo{volume}{4}}, \bibinfo{pages}{936} (\bibinfo{year}{2008}).

\bibitem{Feng}
\bibinfo{author}{Fang, F.~F.} \& \bibinfo{author}{Stiles, P.~J.}
\newblock \bibinfo{title}{Effects of a tilted magnetic field on a two-dimensional electron gas}.
\newblock \emph{\bibinfo{journal}{Phys. Rev.}} \textbf{\bibinfo{volume}{174}},
  \bibinfo{pages}{823} (\bibinfo{year}{1968}).

\bibitem{CFInTiltedField}
\bibinfo{author}{Gee, P.~J.} \emph{et~al.}
\newblock \bibinfo{title}{Composite fermions in tilted magnetic fields and the effect of the confining potential width on the composite-fermion effective mass}.
\newblock \emph{\bibinfo{journal}{Phys. Rev. B}} \textbf{\bibinfo{volume}{54}},
  \bibinfo{pages}{R14313--R14316} (\bibinfo{year}{1996}).

\bibitem{DasSarmaWaveFunctionWidth}
\bibinfo{author}{Zhang, F.~C.} \& \bibinfo{author}{Das~Sarma, S.}
\newblock \bibinfo{title}{Excitation gap in the fractional quantum Hall effect: Finite layer thickness corrections}.
\newblock \emph{\bibinfo{journal}{Phys. Rev. B}} \textbf{\bibinfo{volume}{33}},
  \bibinfo{pages}{2903--2905} (\bibinfo{year}{1986}).

\bibitem{HalperinNelson1978}
\bibinfo{author}{Halperin, B.~I.} \& \bibinfo{author}{Nelson, D.~R.}
\newblock \bibinfo{title}{Theory of two-dimensional melting}.
\newblock \emph{\bibinfo{journal}{Phys. Rev. Lett.}}
  \textbf{\bibinfo{volume}{41}}, \bibinfo{pages}{121--124}
  (\bibinfo{year}{1978}).

\bibitem{Morf1979}
\bibinfo{author}{Morf, R.~H.}
\newblock \bibinfo{title}{Temperature dependence of the shear modulus and  melting of the two-dimensional electron solid}.
\newblock \emph{\bibinfo{journal}{Phys. Rev. Lett.}}
  \textbf{\bibinfo{volume}{43}}, \bibinfo{pages}{931--935}
  (\bibinfo{year}{1979}).

\bibitem{Young1979}
\bibinfo{author}{Young, A.~P.}
\newblock \bibinfo{title}{Melting and the vector coulomb gas in two dimensions}.
\newblock \emph{\bibinfo{journal}{Phys. Rev. B}} \textbf{\bibinfo{volume}{19}},
  \bibinfo{pages}{1855--1866} (\bibinfo{year}{1979}).

\bibitem{Strandburg1988}
\bibinfo{author}{Strandburg, K.~J.}
\newblock \bibinfo{title}{Two-dimensional melting}.
\newblock \emph{\bibinfo{journal}{Rev. Mod. Phys.}}
  \textbf{\bibinfo{volume}{60}}, \bibinfo{pages}{161--207}
  (\bibinfo{year}{1988}).

\bibitem{ME1-PRB2004}
\bibinfo{author}{Spivak, B.} \& \bibinfo{author}{Kivelson, S.~A.}
\newblock \bibinfo{title}{Phases intermediate between a two-dimensional  electron liquid and Wigner crystal}.
\newblock \emph{\bibinfo{journal}{Phys. Rev. B}} \textbf{\bibinfo{volume}{70}},
  \bibinfo{pages}{155114} (\bibinfo{year}{2004}).

\bibitem{ME3-RMP2010}
\bibinfo{author}{Spivak, B.}, \bibinfo{author}{Kravchenko, S.~V.},
  \bibinfo{author}{Kivelson, S.~A.} \& \bibinfo{author}{Gao, X. P.~A.}
\newblock \bibinfo{title}{Colloquium:transport in strongly correlated two-dimensional electron fluids}.
\newblock \emph{\bibinfo{journal}{Rev. Mod. Phys.}}
  \textbf{\bibinfo{volume}{82}}, \bibinfo{pages}{1743--1766}
  (\bibinfo{year}{2010}).

\bibitem{CFModel1}
\bibinfo{author}{Geraedts, S.~D.} \emph{et~al.}
\newblock \bibinfo{title}{The half-filled Landau level: The case for Dirac composite fermions}.
\newblock \emph{\bibinfo{journal}{Science}} \textbf{\bibinfo{volume}{352}},
  \bibinfo{pages}{197} (\bibinfo{year}{2016}).

\bibitem{CFModel2}
\bibinfo{author}{Wang, C.} \& \bibinfo{author}{Senthil, T.}
\newblock \bibinfo{title}{Composite fermi liquids in the lowest Landau level}.
\newblock \emph{\bibinfo{journal}{Phys. Rev. B}} \textbf{\bibinfo{volume}{94}},
  \bibinfo{pages}{245107} (\bibinfo{year}{2016}).

\bibitem{CFModel3}
\bibinfo{author}{Son, D.~T.}
\newblock \bibinfo{title}{Is the composite fermion a dirac particle?}
\newblock \emph{\bibinfo{journal}{Phys. Rev. X}} \textbf{\bibinfo{volume}{5}},
  \bibinfo{pages}{031027} (\bibinfo{year}{2015}).

\bibitem{CFModel4}
\bibinfo{author}{Wang, C.} \& \bibinfo{author}{Senthil, T.}
\newblock \bibinfo{title}{Half-filled landau level, topological insulator surfaces, and three-dimensional quantum spin liquids}.
\newblock \emph{\bibinfo{journal}{Phys. Rev. B}} \textbf{\bibinfo{volume}{93}},
  \bibinfo{pages}{085110} (\bibinfo{year}{2016}).

\end{thebibliography}
\end{document}